\documentclass[12pt]{llncs}
\usepackage{graphicx}
\usepackage{xcolor}
\usepackage{multirow}
\usepackage{enumitem}
\usepackage{amssymb}
\usepackage{amsmath}
\usepackage{mathtools}

\DeclarePairedDelimiter\floor{\lfloor}{\rfloor}
\begin{document}
\title{A Lightweight, Anonymous and Confidential Genomic Computing for Industrial Scale Deployment}
\author{Huafei Zhu}
\institute{Agency for Science, Technology and Research(A*STAR), Singapore}
\maketitle
\begin{abstract}
In Genomic era, researchers and biologists are exploring how human genes influence health, disease and biological pathways, and how the knowledge gained can contribute to better health through prediction disease risk, prevention and personalize therapies. Genome-Wide Association Study (GWAS) is a mechanism that involves rapidly scanning markers across the complete sets of DNA, or genomes, of human to discover the associations between complex human disease/traits and common genetic variants. The insightful analysis on GWAS demands aggregation of data from multiple genome research communities, or international collaborators. This paper studies anonymous and confidential genomic case and control computing within the federated framework leveraging SPDZ. Our contribution mainly comprises the following three-fold:

\begin{itemize}
\item In the first fold, an efficient construction of Beaver triple generators (BTGs) formalized in the 3-party computation leveraging multiplicatively homomorphic key management protocols (mHKMs) is presented and analysed. Interestingly, we are able to show the equivalence between BTGs and mHKMs. We then propose a lightweight construction of BTGs, and show that our construction is secure against semi-honest adversary if the underlying multiplicatively homomorphic encryption is semantically secure.

\item In the second fold, a decoupling model for SPDZ with explicit separation of BTGs from MPC servers (MPCs) is introduced and formalized, where BTGs aim to generate the Beaver triples while MPCs to process the input data. A new notion, which we call blind triple dispensation protocol, is then introduced for securely dispensing the generated Beaver triples, and constructed from mHKMs. We demonstrate the power of mHKMs by showing that it is a useful notion not only for generating Beaver triples but also for securely dispensing triples as well.  

\item In the third-fold, a lightweight genomic case and control computing model is proposed, which reaches the anonymity and confidentiality simultaneously. An efficient truncation algorithm leveraging the depicted BTGs above is then proposed by eliminating computational cost heavy PRandBitL() and PRandInt() protocols involved in the state-of-the-art solutions and thus largely benefits us computing residual vectors for industrial scale deployment.

\end{itemize}
\end{abstract}

\section{Introduction}
Genome-wide association study (GWAS) is a mechanism that involves rapidly scanning markers across large volume of genomes to discover the associations between complex human traits and common genetic variants. The insightful analysis on GWAS demands aggregation of data from multiple genome research communities and the free access to inherently sensitive sequenced genomes are required to improve the power of GWAS in identifying disease-causing genetic variants and to help in the discovery of new drug targets. As such, the key problem is a need of trusted environment for federated analysis. Since genomic sequence data such as samples and single nucleotide polymorphisms (SNPs) carry rich personal information, it is paramount importance to protect identities of individuals in the shared data~\cite{Luca2020,Marcelo2020,Marcon2021,Jones2021}. While the $k$-anonymity is a well-known solution which has been extensively deployed in the real world~\cite{Sweeney02}, there is a remarkable recent result by Artomov et al~\cite{Artomov2018} who have proposed an interesting platform for case-control matching, where the singular value decomposition methodology is deployed to express the original matrix of genotypes as a multiplication of three matrices. Namely, a case genotype matrix G (=$USV^T$) is firstly decomposed into $U$ (a basis of principal directions with the first direction chosen along the largest variance in the original data), $S$ (a diagonal matrix of singular values) and $V$ (a matrix of principal components that provides coordinates of original samples in the basis of matrix $U$); the left singular vectors matrix $U$ is used to reconstruct an approximation matrix for genomic case dataset with the help of statistics information (say, the number of reference alleles, homozygous alleles and heterozygous alleles). From the anonymity point of view, the data owner can share $U$ unrestrictedly with remote database/control server since it does not contain any information on the individual level data. 

\subsection{The motivation of this work}
For a given left singular matrix $U$ which is shared by case data owner Alice, the control data owner Bob, can compute an approximation of $U$ leveraging the residual vector computing technique. That is, Bob first computes projection norm $l$ (=: $||(I- UU^T)z||$) which is compared with a predefined threshold $\tau$ (reject if and only if $l < \tau$); then genomic inflation factor $\lambda$ is calculated to check whether the Hardy-Weinberg equilibrium (HWE) is satisfied. Namely, on input case and current control summary statistical information, Bob computes genomic inflation factor $\lambda$ which in turn is computed from the ration of Allelic chi-test and Armitage trend for maintaining the control quality. More precisely, for a given genotype, we assume that Ref is a recessive allele while Alt is a dominant allele. Then one defines an encoding strategy $E$ such that $E$(Ref,Ref) =0, $E$(Ref, Alt) =1 and $E$(Alt, Alt) =2

\begin{table}

\label{table}
\setlength{\tabcolsep}{4pt}

\centering
\begin{tabular}{|p{35pt}|p{35pt}|p{35pt}|p{35pt}|p{35pt}|}
\hline

       & 0 & 1 & 2 & Total  \\

\hline
Case   & $r_0$ & $r_1$ & $r_2$ & R  \\

Control & $s_0$ & $s_1$ & $s_2$ & S  \\

Total & $n_0$ & $n_1$ & $n_2$ & N  \\
\hline
\end{tabular} 
\label{table:Single SNP and multiple samples}
\medskip
\caption{Genotype distribution - (Ref, Alt)}
\end{table}
Let $X_G ^2$ be the Chi-square test for association and $X_A ^2$ be the Armitage's trend test. It is well known to get good quality of controls, $\frac{X_A ^2}{X_G ^2}$ approximates 1. 
$$\frac{X_A ^2}{X_G ^2} = 1 + \frac{4n_0 n_2 - n_1 ^2}{(n_1 + 2n_2)(n_1 + 2n_0)}$$
For example, in Table.1, we have $r_0$ (Ref, Ref) allelic pairs in the case set and  $s_0$ allelic pairs in the control set. Let $X_G ^2$ be the Chi-square test for association and $X_A ^2$ be the Armitage's trend test. It is shown in~\cite{Peter1997} the fact that $\frac{X_A ^2}{X_G ^2} = 1 + \frac{4n_0 n_2 - n_1 ^2}{(n_1 + 2n_2)(n_1 + 2n_0)}$. A derived control genotype dataset reaches controlled equality if and only if the Hardy-Weinberg equilibrium roughly holds in the combined samples where $\lambda $ (:= $\frac{4n_0 n_2 - n_1 ^2}{(n_1 + 2n_2)(n_1 + 2n_0)}$) is approximately equal to zero. This test benefits researchers weeding out variables that are either ill-conditioned or simply contain no information that will help us predict the action of interest. We refer to the reader~\cite{Peter1997,cho18} for more details. 

In summary, to reach the anonymity of GWAS case and control computing, the singular value decomposition methodology is a prospective solution. To reach the confidentiality, the computed left singular matrix $U$ as well as the statistics information $(r_0, r_1, r_2)$ related to the case dataset should be well protected. Although, there are solutions to reach the anonymity and confidentiality separately in the context of GWAS case and control computing (say~\cite{Marcelo2020,Artomov2018,cho18} and the reference therein), it is a challenging to reach the anonymity and confidentiality of genomic datasets simultaneously.

\subsubsection{What's not addressed in this paper}
For a given genomic case dataset that includes samples, SNPs (or loci of SNPs) and encoding strategy $E$ for (Ref,Ref), (Ref, Alt) and (Alt, Alt), the data owner Alice can define a genomic case matrix G. There are known methods such as the singular vector decomposition, principal component analysis and iterative eigenvalue techniques for Alice to compute the normalized left singular matrix $U$ (i.e., $U_{m \times n} U^T $ = $I_{m \times m}$) efficiently. We view $U$ as a private data silo in the context of federated computing where all computations over private silos are processed on-premise. Hence we do not address how to compute $U$ and statistics information $(r_0, r_1, r_2)$ of genomic case datasets.      

\subsubsection{What are focused in this paper}
Considering a scenario where Alice holds a private genomic case dataset case(DB) (together with her private data $U$ and $(r_0, r_1, r_2)$ while Bob holds private genomic control dataset ctrl(DB) and its statistics information $(s_0, s_1, s_1)$. What we are focusing in this paper is \emph{how to efficiently compute residual vectors \rm{ctrl}$(U)$ $\stackrel{def}{=}$ $\{z \in \rm{ctrl(DB)}$: $||(I- UU^T)z|| \leq \tau\}$ such that the inflation factor $\lambda$ defined over $\rm{case(DB)}$ $\cup$ \rm{ctrl}$(U)$ is approximately equal to zero (say, $\tau $= 0.3 and $\lambda$ $\leq $ 0.05).}

\subsection{The challenging problem} 
To securely compute residual vectors with controlled inflation factors, one may first call linear computing protocols for fixed-point data format. Then, a comparison protocol is called to check whether the value $||(I- UU^T)z||$ and the inflation factor $\lambda$ computed are within the scope of the predefined thresholds. Our tool to solve the problem is leveraging the concept of secure multi-party computation which is well studied in cryptography~\cite{Goldreichbook1,Goldreichbook2}. Numerous good solutions are available such as the ring-based zero-splitting~\cite{Dan2012,aby15,Dan2016,Lindell2016} and SPDZ solutions~\cite{Ivan2012,Ivan2013,Ivan2018,Smart1901,Smart2001}. Each of solutions has its own pros and cons. For example, although the ring-based solutions (say, Sharemind) is efficient, the scalability to support multi-parties is problematic. If Sharemind works in the federated learning architecture, then each data provider plays the role of MPC server. Assuming that an application is to aggregate data from more than three silos, then secure computation protocols involving more than 3-party are highly demanded. On the other hand, while SPDZ provides high scalability, the efficiency of the underlying  cryptographic solutions to generate Beaver triples is a challenging task. 

To the best of our knowledge, the most efficient solution working in the SPDZ framework is MASCOT where the notion of oblivious transfer extension (OT-extension) has been applied successfully for generating message authentication codes (MACs)~\cite{Keller16}. The number of invoked OTs is the size of the underlying finite field $F$ where a secret sharing of multiplication $k \times m$ = $a + b \in F$ is performed, $k \in F$ is a private authentication key of Alice and $m \in F$ is a private message of Bob; the output of Alice is $a \in F$ and the output of Bob is $b \in F$. To attain an acceptable security, MASCOT assumes that the size of field is 128-bit and hence 128 OTs will be invoked for each computation (we assume that efficient OTs~\cite{Orlandi15} are applied here). We emphasize that the OT-extension technique applies to the case where the private input $k \in F$ is fixed. We do not know how to apply the OT-extension technique for general multiplicative computations within the MASCOT framework where both $k \in F$ and $m \in F$ are random values. 

Besides the challenging task for constructing efficient Beaver triple generators, we know that efficient solutions to truncation protocols that will be used to process the fixed-point data format and comparison protocols that will be used to maintain the quality of genomic control datasets are highly demanded. Notice that both truncation and comparison (or sign) protocols can be constructed from secure linear computation protocols which in turn, can be reduced to efficient solutions to the multiplications within the SPDZ framework. An interesting yet challenging problem is thus $-$ \emph{how to construct efficient Beaver triple generators leveraging flexible cryptographic primitives beyond the state-of-the-art somewhat fully homomorphic encryption and OT-extension based solutions for industrial scale deployment?} 

\subsection{The techniques}
In~\cite{cho18}, Cho et al proposed a single/centric solution for constructing Beaver triple generators. In their model, a trusted Beaver triple generator produces $a, b, c \in Z^*_p$ such that $c =ab$~mod~$p$ and then dispenses the shares ($[a]$, $[b]$ and $[c]$) of each value to the corresponding MPC servers. In~\cite{zhu18}, an interesting notion called multiplicatively homomorphic key management (mHKM) protocol is introduced and formalized in the context of multi-party computation, which in essence is an instance of general concept of transitive computing~\cite{zhu19}. In their model, a data owner $O$ and a set of key management(KM) servers $S_1, \cdots, S_n$ are introduced to collaboratively dispense the master key $k$ managed by the data owner $O$. A KM scheme is multiplicatively homomorphic if $k$ =$k_O k_1 \cdots k_n$, where multiplicatively homomorphic encryptions are performed over the underlying cyclic group. The definition of $k_O$ benefits $O$ controlling the dependency of shares $(k_1, \cdots, k_n)$ for cloud key management. If we remove $k_O$ and restrict $n$ =2, then a secure computation for the relationship $k$ =$k_1 \times k_2$ can be derived trivially which in essence, is an instance of Beaver triples, where $k_i$ is privately computed by each of Beaver triple generators BTG$_i$ ($i \in {A, B}$) and $k$ by BTG$_C$~\cite{zhu20}. In this paper, we are able to show that two interesting notions BTGs and mHKMs are equivalent. That is, given an mHKM protocol, one can efficiently construct BTGs and vice versa. We further show that BTGs can be used to blindly dispense Beaver triples as well. Consequently, mHKMs are backbones not only for constructions of BTGs but also useful for secure triple dispensation.    

The remaining question now will be how to dispense shares of each $k_i$ and $k$ to MPC servers without involving somewhat homomorphic encryptions or oblivious transfer protocols, from which, efficient arithmetic and non-arithmetic protocols such as truncation, sign (comparison) protocols and more can be constructed. To this ends, we introduce a new notion called blind triple dispensation, where a secure triple dispensation mechanism is implemented by secure Beaver triple generators. This interesting result shows that the notion of mHKM can be deployed not only for generating Beaver triples but also for securely dispensing the generated Beaver triples as well.  

\subsection{Our contribution}
In this paper, an efficient construction of Beaver triple generators (BTGs) leveraging multiplicatively homomorphic key management (mHKM) system is presented and analysed. The idea behind our construction is that we first define a 3-party mHKM protocol, and then map this 3-party mHKM protocol to a 3-party BTGs protocol, where each independent BTG generates its private value (say, BTG$_A$ and BTG$_B$ generating $a$ and $b$ respectively) and a dependent BTG (say, BTG$_C$) privately computes its value $c$ from secretly encoded values $a$ and $b$ such that $c = ab$. We are able to show that if the underlying mHKM protocol is semantically secure, then the proposed construction of BTGs is secure against semi-honest adversaries. 

We further introduce decoupling model for secure genomic case and control computation for industrial scale deployment. In this paper, two new solutions are presented for  data owners blindly dispensing the generated Beaver triples. The decoupling model, together with the introduced blind triple dispensation, benefits us constructing lightweight BTGs for industrial scale deployment leveraging efficient, flexible cryptographic primitives beyond the state-of-the-art somewhat fully homomorphic encryption and OT-extension based solutions. 

Finally, a lightweight anonymous and confidential genomic case and control computing model is proposed. An efficient truncation algorithm leveraging BTGs depicted above is then proposed by eliminating computational cost heavy PRandBitL() and PRandInt() protocols involved in the state-of-the-art solutions by Catrina et al~\cite{Catrina10A,Catrina10B} and thus benefits us computing residual vectors for industrial scale deployment. 

\subsubsection{The roadmap} The rest of this paper is organized as follows: In section 2, notions and notations of MPC is sketched that will be used to prove the equivalence between mHKMs and BTGs in Section 3. We provide blind triple dispensation protocol in Section 4, and an efficient implementation of secure genomic case and control computing in Section 5. We conclude our work in Section~6.

\section{SMPC: notions and notations}

\subsubsection{Secure multi-party computation} 
A multi-party computation securely computes an $m$-ary functionality (i.e., secure multi-party computation, SMPC) if the following simulation-based security definition is satisfied~\cite{Goldreichbook2}.

Let $[m]$ = $\{1$, $\cdots$, $m\}$. For $I \in \{i_1, \cdots, i_t\}$ $\subseteq$ $[m]$, we let $f_I(x_1, \cdots, x_m)$ denote the subsequence $f_{i_1}(x_1, \cdots, x_m)$, $\cdots$, $f_{i_t}(x_1, \cdots, x_m)$. Let $\mathrm{\Pi}$ be an $m$-party protocol for computing $f$. The view of the $i$-th party during an execution of $\mathrm{\Pi}$ on $\overline{x}$:= $(x_1, \cdots, x_m)$ is denoted by $\mathrm{View_i ^{\Pi}} (\overline{x})$. For $I$ = $\{i_1, \cdots, i_t\}$, we let $\mathrm{View_I ^{\Pi}} (\overline{x})$:= ($I$, $\mathrm{View_{i_1} ^{\Pi}} (\overline{x})$, $\cdots$, $\mathrm{View_{i_t} ^{\Pi}} (\overline{x})$). 

\begin{itemize}
\item In case $f$ is a deterministic $m$-ary functionality, we say $\mathrm{\Pi}$ privately computes $f$ if there exists a probabilistic polynomial-time algorithm denoted $S$, such that for every $I \subseteq [m]$, it holds that $S(I$, $(x_{i_1}, \cdots, x_{i_t})$, $f_I(\overline{x}))$ is computationally indistinguishable with $\mathrm{View_I ^{\Pi}} (\overline{x})$. 

\item In general case, we say $\mathrm{\Pi}$ privately computes $f$ if there exists a probabilistic polynomial-time algorithm denoted $S$, such that for every $I \subseteq [m]$, it holds that $S(I$, $(x_{i_1}, \cdots, x_{i_t})$, $f_I(\overline{x})$, $f(\overline{x}))$ is computationally indistinguishable with $\mathrm{View_I ^{\Pi}}$ $((\overline{x})$, $f(\overline{x}))$.
\end{itemize}

\subsubsection{Oracle-aided multi-party computation} 
An oracle-aided protocol is a protocol augmented by a pair of oracle types, per each party. An oracle-call step is defined as follows: a party writes an oracle request on its own oracle tape and then sends it to the other parties; in response, each of the other parties writes its query on its own oracle tape and responds to the first party with an oracle call message; at this point the oracle is invoked and the oracle answer is written by the oracle on the ready-only oracle tape of each party. An oracle-aided protocol is said to privately reduce $g$ to $f$ if it securely computes $g$ when using the oracle-functionality $f$. In such a case, we say that $g$ is securely reducible to $f$.

\section{On the equivalence between BTGs and mHKMs}
In this section, we are able to show that two interesting notions BTGs and mHKMs are equivalent. That is, given an mHKM protocol, we can construct BTGs and vice versa. In the next section, we will further show that BTGs can be deployed to blindly dispense the generated Beaver triples as well. Consequently, mHKMs are backbone not only for constructions of BTGs but also powerful tools for secure triple dispensations.   
  
\subsection{Beaver triple functionality}
We write $[x]$ to mean that each party $P_i$ holds a random, additive sharing $x_i$ of $x$ such that $x$ = $x_1 + \cdots + x_n$, where $i=1, \cdots, n$. The values are stored in the dictionary Val defined in the functionality $\mathcal{F}_{\rm {Triple}} $~\cite{Keller16}. Please refer to the Table.2 for more details.

\begin{table}
\label{table}
\noindent \framebox{\parbox[c]{16cm}{\center{\underline{The functionality of Beaver triple generator $\mathcal{F}_{\rm {BTG}} $}} 
\begin{itemize}
\item The functionality maintains a dictionary, Val to keep track of assigned value, where entry of Val lies in a fixed field. 
\item On input (Triple, id$_A$,id$_B$, id$_C$) from all parties, sample two random values $a$, $b$ $\leftarrow$ $F$, and set [ Val[id$_A$], Val[id$_B$], Val[id$_C$] ] $\leftarrow $ ($a$, $b$, $ab$).
\end{itemize}
}}
\medskip
\caption{Beaver triple functionality}
\end{table}

\subsection{Syntax of multiplicatively homomorphic key management protocol}
Intuitively, a multiplicatively homomorphic key management (mHKM) protocol is a key dispensation procedure based on multiplicatively homomorphic encryption~\cite{ELG85}. An mHKM protocol comprises a data owner and a set of application servers together with a database server. 

\subsubsection{Multiplicatively homomorphic key management functionality}
Each party $P_i$ holds a random, multiplicative sharing $x_i$ of $x$ such that $x$ = $x_1  \cdots  x_n$, where $i=1, \cdots, n$. The values are stored in the dictionary Val defined in the functionality $\mathcal{F}_{\rm {mHKM}}$. Please refer to the Table.3 for more details.

\begin{table}
\label{table}
\noindent \framebox{\parbox[c]{16cm}{\center{\underline{The functionality of multiplicatively homomorphic key management $\mathcal{F}_{\rm {mHKM}} $}} 
\begin{itemize}
\item The functionality maintains a dictionary, Val to keep track of assigned value, where entry of Val lies in a fixed field. 
\item On input (mHKM, id$_1$, $\cdots$, id$_{n-1}$, id$_{n}$) from all parties, sample random values $a_1$, $\cdots$, $a_{n-1}$ $\leftarrow$ $F$, and set $a_n$ $\leftarrow$ $a_1\cdots a_{n-1}$, [Val[id$_1$], $\cdots$, Val[id$_n$] ] $\leftarrow $ ($a_1$, $\cdots$, $a_n$).
\end{itemize}
}}
\medskip
\caption{A description of multiplicatively homomorphic key management functionality}
\end{table}

\begin{definition}
(~\cite{zhu18}) A key management(dispensation) protocol is a multi-party computation consisting of a data owner $O$ and a set of servers $S_1, \cdots, S_n$ who collaborate with each other to dispense master key $k$ by running the following algorithms:
\begin{itemize}
\item key generation algorithm: on input a security parameter $1^{\kappa}$, it generates a public and secret key pair for participants:
\begin{itemize}
\item on input a security parameter $1^{\kappa}$, an instance of semantically secure encryption scheme is invoked. The public key of instance is denoted by $pk_E$ (notice that the secret $sk_E$ is not an output; Hence our protocol is defined in the common public key model);

\item on input $pk_E$, a data owner generates a pair of public and secret key pair $(pk_O, sk_O)$;

\item on input $pk_E$, individual server $S_i$ generates a pair of public and secret key pair $(pk_i, sk_i)$ independently, $i=1, \cdots, n$.
\end{itemize}

\item On input $(pk_E, pk_O, pk_i)$, and a common reference public key $pk$ that will be used to dispense the master key of the data owner is generated, $i=1, \cdots, n$.

\item on input a secret key $k$ and $pk$, a multi-party processing is running among the data owner and the $n$ servers; The output of each server is a random key $k_i$ that is secretly selected by each server $S_i$.
\end{itemize}

We say, a key management scheme is multiplicatively homomorphic if $k$ =$k_O k_1 \cdots k_n$ (a multiplication procedure is performed over a cyclic group).
\end{definition}

\subsection{A construction of BTG based on mHKM}
In this section, a new construction of BTGs is proposed which is different from the known result~\cite{zhu20}. The solution presented in~\cite{zhu20} starts from secure computations of $c$ and $a$ and then $b = c a^{-1}$. In this paper, we propose a concise procedure to compute Beaver triple by starting computations of $a$ and $b$ and then $c =ab$. Consequently, the inversion procedure required in~\cite{zhu20} can be eliminated.

Let $p$ be a large safe prime number, $p$ =2$q$ +1, $p$ and $q$ are prime numbers. Let $G \subseteq Z_p ^*$ be a cyclic group of order $q$ and $g$ be a generator of $G$. Let $h_i$ = $g^{x_i}$ mod $p$, where $x_i \in_R [1, q]$ is randomly generated by the Beaver multiplication triple generator BTG$_i$ ($i \in \{A, B, C\}$). 

\begin{table}
\label{table}
\noindent \framebox{\parbox[c]{16cm}{\center{BTGs based on mHKM protocols} 
\begin{itemize}
\item Step 1. BTG$_A$ performs the following computations: 
\begin{itemize}
\item computing $h_{AC}$ = $h_A \times h_C$~mod~$p$; 
\item randomly selecting $r_A \in Z_q$, $a \in Z^* _p$ and then computing $u_A$ = $g^{r_A}$~mod~$p$ and $v_A$ = $a \times h_{AC} ^{r_A}$~mod~$p$; 
\item sending $(u_A, v_A)$ to BTG$_B$;
\end{itemize}

\item Step 2. upon receiving $(u_A, v_A)$ from BTG$_A$, BTG$_B$ performs the following computations:
\begin{itemize}
\item randomly selecting $r_B \in Z_q$ $, b \in Z^* _p$  and computing $u_B$ =  $u_A g^{r_B}$ and $v_B$ = $v_A \times b \times h_{AC} ^{r_B}$~mod~$p$ (= $a\times b \times h_{AC} ^{r_A} \times h_{AC} ^{r_B}$~mod~$p$; 

\item sending $(u_B, v_B)$ to BTG$_A$
\end{itemize}

\item Step 3: upon receiving $(u_B, v_B)$ from BTG$_B$, BTG$_A$ performs the following computations:
\begin{itemize}
\item $v_C$ $\stackrel{def}{=}$ $v_B /{u_B ^{x_A}}$ (= $ a\times b \times h_{AC} ^{r_A} \times h_{AC} ^{r_B}/{u_B ^{x_A}}$ = $a \times b \times (h_A \times h_C)^{(r_A + r_B)}/{u_B^{x_A}}$ = $a\times b \times h_C ^{(r_A + r_B)})$;

\item computing $u_C$ $\stackrel{def}{=}$ $u_B \times g^{r'_B}$ (= $g^{(r_A + r_B + r'_B)}$~mod~$p$) and $v_C$ $\leftarrow v_C \times h_C ^{r'_B}$~mod~$p$, and then sending $(u_C, v_C)$ to BTG$_C$.

\end{itemize}

\item upon receiving $(u_C, v_C)$, BTG$_A$ performs the following computations:
\begin{itemize}
\item computing $c$ $\stackrel{def}{=}$ $ v_C/{u_C ^ {x_C}}$~mod~$p$.
\end{itemize}
\end{itemize}
}}

\medskip
\caption{A construction of BTG protocol based on mHKM}
\label{Fig.1}
\end{table}

By applying the same proof technique~\cite{zhu20} and assuming that the underlying El Gamal encryption is semantically secure, we claim that the BTG protocol described above is secure against semi-honest adversary.

\subsection{A construction of mHKM based on BTG}
In this section, we provide a construction of multiplicatively homomorphic key management protocols based on the Beaver triple generators depicted above. The idea behind of our construction is that $-$ for each group of triple generators (BTG$_n$, BTG$_{n-1}$, BTG$'_{n-2}$), we generate $a_n$ = $a_{n-1}a'_{n-2}$. This $a'_{n-2}$ will be viewed as a master key and the owner of this value invokes an instance of BTGs as a party computation (BTG$'_{n-2}$, BTG$_{n-2}$, BTG$'_{n-3}$) to generate a triple $(a'_{n-2}, a_{n-2}, a'_{n-3})$ such that $a'_{n-2}$ =$a_{n-2} a'_{n-3}$. The procedure continues until we get $a_n$ =$a_{n-1} \cdots a_1$. The details are depicted in Table.5. One can verify that from (BTG$_n$, BTG$_{n-1}$, BTG$'_{n-2}$),  (BTG$'_{n-2}$, BTG$_{n-2}$, BTG$'_{n-3}$), $\cdots$, (BTG$_3$, BTG$_2$, BTG$_1$), we know that $a_n$ =$a_{n-1} \cdots a_1$. A rigorous proof of our construction is presented in Theorem 1. Consequently, we are able to show the equivalence between two different cryptographic primitives.

\begin{table}
\noindent \framebox{\parbox[c]{15.8 cm}{\center{\underline{A description of mHKM protocol leveraging BTG}} 
\begin{itemize}
\item Invoking an instance of BTG protocol with three parties BTG$_n$, BTG$_{n-1}$ and BTG$'_{n-2}$ such that $a_n$ =$a_{n-1} \times a' _{n-2}$, assuming that BTG$_i$ generates its private value $a_i$ and relaying BTG$'_i$ generates $a' _i$, ($i$ = $1,\cdots, n$); 
\item Invoking an instance of BTG protocol with three parties BTG$'_{n-2}$, BTG$_{n-2}$, BTG$_{n-3}$ such that $a' _{n-2}$ =$a_{n-2} \times a'_{n-3}$. 
\item The invocation procedure of BTG instances continues until it reaches final result $a'_2 = a_2 \times a' _1$ (:=$ a_2 \times a_1$), where $a_1 \leftarrow a'_1$.
\end{itemize}
}}
\medskip
\caption{A construction of mHKMs based on BTGs}
\end{table}

\subsection{The proof of security}

\begin{theorem}
Let $\mathrm{g_{mHKM}}$ be the multiplicatively homomorphic key management functionality. Let $\mathrm{\Pi^{g_{mHKM}|f_{BTG}}}$ be an oracle-aided protocol that privately reduces $\mathrm{g_{mHKM}}$ to $\mathrm{f_{BTG}}$ and $\mathrm{\Pi^{f_{BTG}}}$ be a protocol privately computes $\mathrm{f_{BGT}}$. Suppose $\mathrm{g_{mHKM}}$ is privately reducible to $\mathrm{f_{BTG}}$ and that there exists a protocol privately computing $\mathrm{f_{BTG}}$, then there exists a protocol for privately computing $\mathrm{g_{mHKM}}$.
\end{theorem}

\begin{proof}
We construct a protocol $\mathrm{\Pi}$ for computing $\mathrm{g_{mHKM}}$ by replacing each invocation of the oracle $\mathrm{f_{BTG}}$ with an execution of protocol $\mathrm{\Pi^{f_{BTG}}}$. Note that in the semi-honest model, the steps executed $\mathrm{\Pi^{g_{mHKM}|f_{BTG}}}$ inside $\mathrm{\Pi}$ are independent the actual execution of $\mathrm{\Pi^{f_{BTG}}}$ and depend only on the output of $\mathrm{\Pi^{f_{BTG}}}$. 

Let $\mathrm{S_i ^{g_{mHKM}|f_{BTG}}}$ and $\mathrm{S_i ^{f_{BTG}}}$ be the corresponding simulators for the view of party $P_i$ (either BTG$_i$ or BTG'$_i$). We construct a simulator $S_i$ for the view of party $P_i$ in $\mathrm{\Pi}$. That is, we first run $\mathrm{S_i ^{g_{mHKM}|f_{BTG}}}$ and obtain the simulated view of party $P_i$ in $\mathrm{{\Pi}^{g_{mHKM}|f_{BTG}}}$. This simulated view includes queries made by $P_i$ and the corresponding answers from the oracle. Invoking $\mathrm{S_i ^{f_{BTG}}}$ on each of partial query-answer $(q_i, a_i)$, we fill in the view of party $P_i$ for each of these interactions of $\mathrm{S_i ^{f_{BTG}}}$. The rest of the proof is to show that $S_i$ indeed generates a distribution that is indistinguishable from the view of $P_i$ in an actual execution of $\mathrm{\Pi}$. 

Let $\mathrm{H_i}$ be a hybrid distribution represents the view of $P_i$ in an execution of $\mathrm{{\Pi}^{g_{mHKM}|f_{BTG}}}$ that is augmented by the corresponding invocation of $\mathrm{S_i ^{f_{BTG}}}$. That is, for each query-answer pair $(q_i, a_i)$, we augment its view with $\mathrm{S_i ^{f_{BTG}}}$. It follows that $\mathrm{H_i}$ represents the execution of protocol $\mathrm{\Pi}$ with the exception that $\mathrm{\Pi ^{f_{BTG}}}$ is replaced by simulated transcripts. We are able to show that the following facts, from which our claim is proved.

\begin{itemize}
\item the distribution between $\mathrm{H_i}$ and $\mathrm{\Pi}$ are computationally indistinguishable: notice that the distributions of $\mathrm{H_i}$ and $\mathrm{\Pi}$ differ $\mathrm{\Pi ^{f_{BTG}}}$ and $\mathrm{S_i ^{f_{BTG}}}$ which is computationally indistinguishable assuming that $\mathrm{\Pi ^{f_{BTG}}}$ securely computes $\mathrm{f_{BTG}}$. 
 
\item the distribution between $\mathrm{H_i}$ and $\mathrm{S_i}$ are computationally indistinguishable: notice that the distributions between ($\mathrm{{\Pi}^{g_{mHKM}|f_{BTG}}}$, $\mathrm{S_i ^{f_{BTG}}}$) is computationally indistinguishable from ($\mathrm{{S_i}^{g_{mHKM}|f_{BTG}}}$, $\mathrm{S_i ^{f_{BTG}}}$). The distribution ($\mathrm{{S_i}^{g_{mHKM}|f_{BTG}}}$, $\mathrm{S_i ^{f_{BTG}}}$) defines $\mathrm{S_i}$. That means $\mathrm{H_i}$ and $\mathrm{S_i}$ are computationally indistinguishable.
\end{itemize}
\end{proof}

\section{Decoupling model for Beaver triple dispensation}
In this section, we provide a decoupling model for Beaver triple dispensation. Different from the single/centric Beaver triple generator case~\cite{cho18}, our model works in the three Beaver triple generators setting, where each of BTGs holds its own private value $a$ or $b$ or $c$ such that $c = a \times b$. We assume that there are $m$-MPC servers (MPC$_1$, $\cdots$, MPC$_m$) running in the federated computing architecture, where each of private data silos plays the role of MPC server as well. We separate roles of BTGs from that of MPCs and assume that the role of each BTG is to generate and dispense triples while the role of each MPC is to process data leveraging the available triples. Borrowing the notion of blind signature, we propose a new protocol, which we call blind triple dispensation (BTD). BTD in essence is a randomized processing for Beaver triple dispensation, which benefits MPCs protecting their private data if multiplications are processed in the framework of SPDZ. The details are below.

\subsection{Blind triple dispensation}
To multiply $x \in F$ and $y \in F$, the data owner $O_x$ of $x$ calls for the Beaver triple component $a$ while data owner $O_y$ of $y$ calls for the Beaver triple component $b$. A splitting of the private value $a$ is defined by $[a]$ = $(a_1, \cdots, a_n)$ (as usual, a random split of data is called a secret share of that data. The process can be viewed as a keyless encryption). Below, two randomized splitting solutions are provided to dispense the generated Beaver triples.

\subsubsection{Single-randomness solution}
$O_x$ and $O_y$ collaboratively generate a random value $r \in F$, which plays the role of blind factor to protect the called triple components. More precisely, $O_x$ and $O_y$ jointly generate a shared value $r \in F$. $O_x$ then sends a request to BTG$_A$ who holds private triple component $a$ while $O_y$ sends a request to BTG$_B$ who holds private triple component $b$. Let $a'$ = $r \times a$ and $b'$ =$r \times b$, it follows that $[a']$ =$r[a]$ and $[b']$ = $r[b]$. $O_x$ (resp., $O_y$) performs the splitting procedure depicted below.

\begin{itemize}
\item  $O_x$ (for simplicity, we assume that $O_x$ is MPC$_1$) selects $a_2, \cdots, a_m \in Z^*_p$ uniformly at random, and then sends $a_2$ to MPC$_2$, $\cdots$, $a_m$ to MPC$_m$ via private channels shared between $O_x$ and MPC$_j$, $j=1, \cdots, m$;

\item $O_x$ computes $a_1$ = $r a$ - $a_2$ - $\cdots$ - $a_m$~mod~p and keeps $a_1$ locally.
\end{itemize}

Since there are $m$-MPC (as above, we simply assume that MPC$_1$ holds $x$ and MPC$_2$ holds private data $y$), an interesting question is who requests $c$. Here we apply the committee selection technique presented in~\cite{Micali18,Micali19} to the set $\{ \rm{MPC}_3, \cdots,\rm{MPC}_n\}$. The selected committee leader (say, MPC$_3$) will make a quest to BTG$_C$ and then computes $[c']$ = $r^2 [c]$, where $r$ is shared with committee leader MPC$_3$. One can verify that $[c']$ = $[a']$ $\times$ $[b']$. We emphasize that the shares of randomized triple ($[a']$, $[b']$, $[c']$) will be used as Beaver triples in the context of SPDZ (instead of $[a]$, $[b]$, $[c]$).  

\subsubsection{Two-randomness solution}: Let $a'$ = $r_a \times a$ and $b'$ =$r_b \times b$, it follows that $[a']$ =$r_a [a]$ and $[b']$ = $r_b [b]$. To compute $r_c$ = $r_a \times r_b$, we can apply three-party computation protocol described in Section~3, where $O_x$, $O_y$ and committee leader (say, MPC$_3$) are involved and the shares $r_a[a]$, $r_b[b]$ and $r_c[c]$ are dispensed respectively.

\subsection{The multiplication}
Assuming that MPC$_1$ holds private data $x$ and MPC$_2$ holds private data $y$. Suppose $m$ parties MPC$_1$, $\cdots$, MPC$_m$ wish to compute $[xy]$ collaboratively given $[x]$ and $[y]$. We assume MPC$_1$ makes a request of $a$ to BTG$_A$, and then applies the depicted blind triple dispensation procedure to get a share $[a]$. Same procedure applies to MPC$_2$ for obtaining a share $[b]$. Borrowing the notation from SPDZ, by $\rho$, we denote an opening of $[x] - [a]$ and by $\epsilon$, an opening of $[y] - [b]$. Given $\rho$ and $\epsilon$, each party can compute his secret share of $[x y]$ = $(\rho + [a])$ $(\epsilon + [b])$ = $[c]$ + $\epsilon [a]$ + $\rho [b]$ + $\rho \epsilon $ locally.

\section{An immediate application to genomic case and control computing}
The computation of residual vector for genomic case and control study is defined over fixed-point data type. Since addition and multiplication operations are defined over integer data type in MPC, we need a truncation protocol bridging the computations between fixed-point data and integer data types.  

\subsection{Secure of inflation factor over integer domain}
As stated in the introduction section, the inflation factor $\lambda $ is defined by $\frac{4n_0 n_2 - n_1 ^2}{(n_1 + 2n_2)(n_1 + 2n_0)}$, where $n_i$ is integer data type and its meaning is explained in Table 1. Notice that $\lambda$ is known to the genomic control dataset owner which allows us to randomize the numerator and denominator of $\lambda$ while the randomization procedure keeps the ratio ($\lambda$) unchanged. The details of computation is depicted below:

\begin{itemize}

\item genomic case and control dataset owners ($O_{\rm{case}}$ and $O_{\rm{ctrl}}$) invoke a mutually agreed $m$-party multiplication protocol to compute a share $[4n_0 n_2 - n_1 ^2]$ of numerator $4n_0 n_2 - n_1 ^2$ and a share $[(n_1 + 2n_2)(n_1 + 2n_0)]$ of denominator $(n_1 + 2n_2)(n_1 + 2n_0)$. This procedure invokes integer addition, multiplication within the SPDZ framework, where all computations are defined over a sufficiently large field $Z_p$. 
\item $O_{\rm{case}}$ randomly selects a value $r_{\rm{case}}$ while $O_{\rm{ctrl}}$ randomly selects a value $r_{\rm{ctrl}}$. $O_{\rm{case}}$ and $O_{\rm{ctrl}}$ collaboratively compute the blind factor $[r_{\rm{case}} \times r_{\rm{ctrl}}]$ which in turn applies to the share of numerator $[4n_0 n_2 - n_1 ^2]$ and denominator $[(n_1 + 2n_2)(n_1 + 2n_0)]$;
\item the final results of randomization of numerator $[[r_{\rm{case}} \times r_{\rm{ctrl}}]$ $\times$ $[4n_0 n_2 - n_1 ^2]]$, and denominator $[[r_{\rm{case}} \times r_{\rm{ctrl}}]$ $\times$ $[(n_1 + 2n_2)(n_1 + 2n_0)]]$ are then opened to $O_{\rm{ctrl}}$ from which $\lambda $ can be computed trivially.   
\end{itemize}

One can verify that the correctness of computation. The security of our computation follows from the composition theorem of multiplication protocol.

\subsection{A lightweight truncation protocol without involving PRandBitL() and PRandInt()}     
The truncation protocol is a backbone for secure computation over fixed-point data type where residual vector is defined. There are known solutions to truncation protocols. Following the interesting work of Catrina et al~\cite{Catrina10A,Catrina10B}, we provide an efficient solution by eliminating the computational cost heavy PRandBitL() and PRandInt() protocols using the decoupling model defined in Section 4. 

Borrowing the notations from\cite{Catrina10A,Catrina10B}, we assume that $k$, $e$, $f$ and $m$ are public integers such that $k>0$, $f \geq 0$, $e$ = $k-f \geq 0$ and $0<m \leq k$. A fixed point number can be written $\tilde{x}$ = $s \dot (d_{e-2}\dots d_0d_{-1}\dots d_{-f})$, where $s \in \{-1, 1\}$, and $e$ is the length of integer part and $f$ is the length of fraction part. Denote $\mathcal{Z}_{<k>}$ = $\{x \in \mathcal{Z}| -2^{k-1} +1 \leq x \leq 2^{k-1} -1\}$ and $\mathcal{Q}_{<k, f>}$ = $\{\tilde{x} \in \mathcal{Q}| \tilde{x} = \bar{x}2^{-f}, \bar{x} \in \mathcal{Z}_k\}$. Signed integers are encoded in $Z_q$ by $x$ $\leftarrow$ $\bar{x}$~mod~$q$, where $\bar{x}$ $\in$ $\mathcal{Z}_{<k>}$ and $q > 2^k$. 

\subsubsection{Truncation protocols without involving PRandBitL() and PRandInt(k)} An input of truncation protocol is $[z]$, where $z \in Z_q$, $\bar{z} \in \mathcal{Z}_{<k>}$ and $z$ = $\bar{z}$~mod~$q$. The output is $[\floor{\frac{\bar{z}}{2^m}}]$. The details are depicted below. 

\begin{table}
\label{table}
\noindent \framebox{\parbox[c]{16cm}{\center{Truncation protocol leveraging blind triple dispensation} 
\begin{enumerate}
\item $[z']$ $\leftarrow$ $2^{k-1} + [z]$. It follows that $z'$ = $2^{k-1} + \bar{z}$~mod~$q$ = $2^{k-1}$ + $\bar{z}$. Consequently, $z'$~mod~$2^m$ = $\bar{z}$~mod~$2^m$;

\item $O_{\rm{case}}$ randomly selects $r' \in [0, 2^m-1]$, and computes $[r']$ over $Z_q$; Similarly, $O_{\rm{ctrl}}$ randomly selects $r'' \in [0, 2^{\kappa + k - m}]$, and then computes shares $[r'']$ over $Z_q$, where $q > 2^{\kappa + k  +1}$. From $[r']$ and $[r'']$, MPC$_i$ can compute its share $[r'']2^m + [r']$ locally over $Z_q$, where MPC$_1$ is $O_{\rm{case}}$, MPC$_2$ is $O_{\rm{ctrl}}$ and committee leader is MPC$_3$;

\item Each MPC$_i$ opens its shares to the selected committee leader MPC$_3$, who, in turn, opens $c$ = $(r + z')$~mod~$q$ = $r + z'$ to all participants;

\item By $c'$ we denote $c$~mod~$2^m$. It follows that $c'$ = $(r''2^m + r' + z' )$~mod~$2^m$ = $r'$ + $z'$~mod~$2^m$ - $u 2^m$, where $u \in \{0, 1\}$ and $u=0$ if $r'$ + $z'$~mod~$2^m$ $<$ $2^m$ and 1, otherwise;

\item Since $\bar{z}$ + $c'$ = $\bar{z}$ + $r'$ + $z'$~mod~$2^m$ -$u 2^m$, it follows that $\bar{z}$ -$z'$~mod~$2^m$ + $u 2^m$ = $\bar{z}$ - $\bar{z}$~mod~$2^m$ + $u 2^m$= $\bar{z}$ + $r'$ -$c'$. Consequently, $\floor*{\frac{\bar{z}}{2^m}} \times 2^m$ = $\bar{z}$ + $r'$ -$c'$. This means that $\floor*{\frac{\bar{z}}{2^m}} \times 2^m$~mod~$q$ = ($\bar{z}$ + $r'$ -$c'$)~mod~$q$. As a result, we know that $[\floor*{\frac{\bar{z}}{2^m}}]$~mod~$q$ = $2^{-m}$ $\times$ ($[z]$ + $[r']$ -$c'$)~mod~$q$.
\end{enumerate}
}}

\medskip
\caption{A construction of truncation protocol without involving PRandBitL() and PRandInt()}
\label{Fig.1}
\end{table}

The proposed truncation protocol is secure against semi-honest adversary since the only message leaked is $c$ during the computation which is guaranteed by the selected security parameter as it is clearly stated in~\cite{Catrina10A,Catrina10B}.

\subsection{Securely computing residual vector}
Recall that to compute qualified residual vectors, we need to check whether a genomic control vector $z$ satisfies the condition $||(I- UU^T)z||$ $\leq$ $\tau$, where $\tau$ is a public or secret value depending on the application scenario. In the following processing, we assume that $\tau$ is a secret value. Let $z'$ = $(I- UU^T)z$, which can be efficiently computed by deploying multiplication and truncation procedures defined in the previous subsections. Furthermore, a sign protocol, or equivalently, a comparison protocol, will be called to test whether the condition $(||z'|| - \tau)< 0$ is satisfied. The state-of-the-art efficient solution without bit decomposition presented in~\cite{Nishide2007,Yang2012} can be applied here. As such, we are able to solve the residual vector computing problem within the proposed decoupling model.

\section{Conclusion}
In this paper, a lightweight, anonymous and confidential genomic case and control computing within the federated SPDZ framework has been proposed. We have solved the residual vector problem by introducing decoupling model to separate BTGs from MPCs which benefits us implementing BTGs efficiently beyond the state-of-the-art somewhat homomorphic encryptions and oblivious transfers. We have shown the equivalence between BTGs and mHKMs and then have presented an efficient construction of BTGs leveraging mHKMs. The introduced blind triple dispensation protocol, which in turn, has been implemented by BTGs, benefits us dispensing the generated Beaver triples securely among MPCs. Finally, a lightweight truncation protocol has been proposed by eliminating computation-cost heavy PRandBitL() and PRandInt() protocols and thus benefits us solving the residual vector problem for industrial scale deployment.

\end{document}